# Autonomous Vehicles for All?

Sakib Mahmud Khan, S.M., Khan

Intelligent Transportation Systems Lead, MITRE Corporation, McLean, VA 22102, USA, sakibkhan@mitre.org

M Sabbir Salek, M.S., Salek*

Ph.D. Candidate, Glenn Department of Civil Engineering, Clemson University, Clemson, SC 29634, USA, msalek@clemson.edu

Vareva Harris, V., Harris

Assistant Vice President, Enrollment Management, Benedict College, Columbia, SC 29204, USA, Vareva.Harris@benedict.edu

Gurcan Comert, G., Comert

Associate Professor, Computer Science, Physics and Engineering Department, Benedict College, Columbia, SC 29204, USA, gurcan.comert@benedict.edu

Eric Morris, E., Morris

Professor, Nieri Department of Construction, Development and Planning, Clemson University, Clemson, SC 29634, USA, emorri7@clemson.edu

Mashrur Chowdhury, M., Chowdhury

Professor, Glenn Department of Civil Engineering, Clemson University, Clemson, SC 29634, USA, mac@clemson.edu

The traditional build-and-expand approach is not a viable solution to keep roadway traffic rolling safely, so technological solutions, such as Autonomous Vehicles (AVs), are favored. AVs have considerable potential to increase the carrying capacity of roads, ameliorate the chore of driving, improve safety, provide mobility for those who cannot drive, and help the environment. However, they also raise concerns over whether they are socially responsible, accounting for issues such as fairness, equity, and transparency. Regulatory bodies have focused on AV safety, cybersecurity, privacy, and legal liability issues, but have failed to adequately address social responsibility. Thus, existing AV developers do not have to embed social responsibility factors in their proprietary technology. Adverse bias may therefore occur in the development and deployment of AV technology. For instance, an artificial intelligence-based pedestrian detection application used in an AV may, in limited lighting conditions, be biased to detect pedestrians who belong to a particular racial demographic more efficiently compared to pedestrians from other racial demographics. Also, AV technologies tend to be costly, with a unique hardware and software setup which may be beyond the reach of lower-income people. In addition, data generated by AVs about their users may be misused by third parties such as corporations, criminals, or even foreign governments. AVs promise to dramatically impact labor markets, as many jobs that involve driving will be made redundant. We argue that the academic institutions, industry, and government agencies overseeing AV development and deployment must act proactively to ensure that AVs serve all and do not increase the digital divide in our society.

---

* Corresponding author.

## 1 WHY AUTONOMOUS VEHICLES NEED TO BE SOCIALLY RESPONSIBLE

In a fair, just, and equitable society, all individuals should enjoy access to essential destinations like work, shopping, education, social activities, and more. Unfortunately, our society has not always made good on this promise. A transportation system based around automobile ownership has made access difficult for those who cannot or will not drive due to age, income, location, disability, and more. Low-density, sprawling land-use patterns make access difficult for those who would prefer to, or have no choice but to, walk, bicycle, or take transit. Low-income travelers tend to make fewer trips and travel fewer miles. Further, the fuel taxes and other fees that fund the transportation system are regressive. Bus transit, on which many low-income persons depend, has been gradually dropping down the priority list in public transit spending.

Today, the development of autonomous vehicles (AVs) promises to reduce vehicle crashes, improve traffic flow and ameliorate congestion, increase the speed and efficiency of goods movement, enhance traveler comfort and well-being, conserve fuel, improve the mobility of those who cannot drive, and more. Some even argue that AVs will lead to an era where car ownership is a thing of the past, and mobility will be rented as a service on a per-trip or monthly basis. However, we cannot take for granted that the automation of the vehicle fleet will be socially responsible in all dimensions.

What social responsibility issues do AVs raise? It is likely that AVs will be more expensive than conventional vehicles, for example, due to the suite of sensors, processing hardware, and software they will need. Thus, AVs may exacerbate social inequality by putting car ownership beyond the means of many low-income community members. Further, low-income persons are far more likely to purchase and drive used cars, and, as the auto fleet typically takes 10-15 years to turn over, it may be a decade or more before automated technology "trickles down" to many low-income drivers.

Also, the transparency of connected AV users' data usage must be maintained; travelers need to be aware of how their data is being used. At the same time, agencies and private firms that collect these data may face ethical issues in terms of how the data are shared and acted upon.

Operators of mobility-as-a-service vehicle fleets may have economic incentives to skimp on service to low-income and rural areas. Also, public agencies must develop frameworks that govern how AVs interact with vulnerable road users such as pedestrians and bicyclists. Another major issue is that AVs may make many jobs, such as driving trucks, buses, and ride-hailing vehicles, obsolete. Since these jobs typically require less formal education, their disappearance may disadvantage working-class people. Thus, to ensure that the benefits of AVs reach all of our society, we need proactive action that meshes AV design and deployment with social responsibility considerations so that AVs reduce, not exacerbate, societal divides.

## 2 STEPS TOWARDS DEVELOPING SOCIALLY RESPONSIBLE AUTONOMOUS VEHICLES

In figure 1, we present the steps in developing socially responsible AVs. First, we need to bring in experts from various disciplines (e.g., engineering, economics, social and behavioral sciences, ethics, law) together to agree upon a social responsibility checklist for AVs that will include requirements such as fairness, equity, and transparency. To ensure AVs promote all the requirements on the social responsibility checklist, and meet all user expectations, diverse and representative data should be collected from field tests and pilot or proof-of-concept deployments. These data will help update and, in some cases, rethink the requirements on social responsibility checklist. These data must be analyzed critically by experts from various disciplines, stakeholders, and policymakers to recognize the current state, scope of improvement, and potential societal impact of AVs, and to help update the social responsibility checklist. Although not comprehensive, the steps presented in figure 1 are indispensable and must be regulated by governing bodies so that they are part of any AV social responsibility framework. In this section, we will discuss social responsibility factors, such as fairness, equity, and transparency, in the context of AV technology.

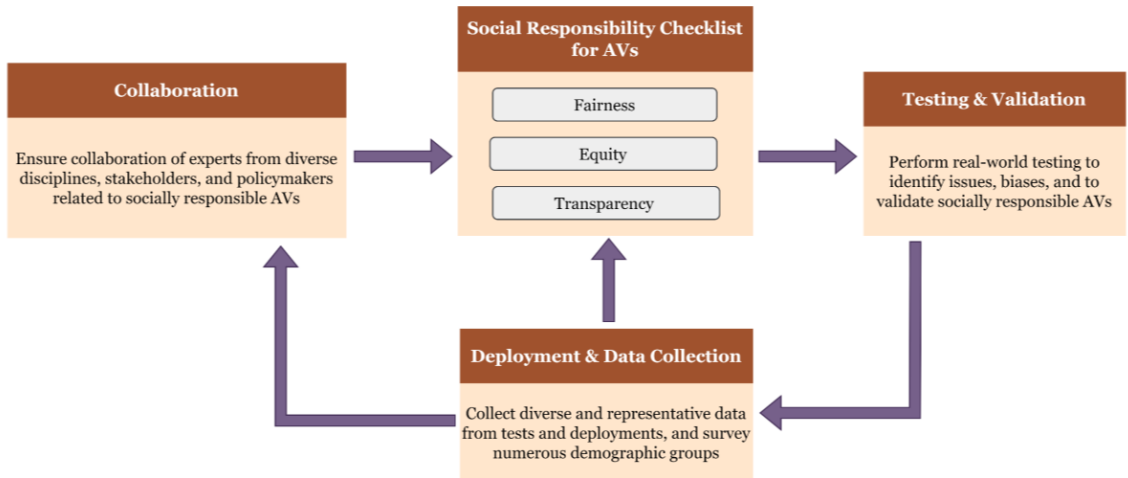

Figure 1: Steps towards developing socially responsible AVs.

**2.1 Fairness in AVs**

Fairness in AVs implies that an AV's underlying decision-making algorithms must not take disparate actions (i.e., beneficial or harmful) for different demographics. In this subsection, we present some algorithmic challenges, fairness metrics, and data challenges for measuring fairness in AVs.

*2.1.1 Algorithmic Challenges*

Developing fair AVs poses algorithmic challenges. Since AVs heavily utilize data-driven AI models, systematic unfairness and biases within the data or the developed model can cause AVs to act unfairly in various situations [7]. AI models utilize training data to make inferences. Thus, if the training data contains biases, these might be replicated by the trained model as well. One way to get rid of such biases is to carefully take out the sensitive attributes or variables that are causing the biases. However, there may remain some attributes or variables that may act as proxies for the already-removed sensitive attributes, and the AI model may pick up the biases from these proxies. Thus, the model itself can act as a source of unfairness.

The rise of generative AI algorithms, such as generative adversarial network (GAN) and generative pre-trained transformer (GPT), has motivated AV industries since breakthroughs in AI have a history of positively influencing AV developments [31]. For example, GANs have been studied for various AV-related applications in recent years including driver behavior imitation [14], sensor modeling [2], and trajectory prediction [22]. On the other hand, being a generative language model, GPT may have some direct impacts, e.g., GPT-based chatbots for infotainment in AVs, as well as some indirect impacts on AV development, e.g., imitation learning-based driver behavior emulation [20,27] may get influenced by GPT's reinforcement learning from human feedback (RLHF) training method [32]. However, generative AI algorithms are also heavily dependent on data and prone to pick up biases from various attributes [23,29]. Therefore, ensuring fairness of these algorithms may pose some additional challenges for AV-related applications.

*2.1.2 Fairness Metrics*

Since fairness in AVs' decision-making depends on the underlying AI algorithms, it is essential to utilize fairness metrics while developing them. To help accurately evaluate the bias present in the AI algorithms, various fairness notions and metrics have been developed for uneven tasks, including group fairness, individual fairness, etc. in the past years.

What does group fairness mean, and how can it help remove biases from AI models used in AVs? The majority of existing group fairness notions are based on the statistical independence between the sensitive information (e.g., race, age, complexion, ability, and gender) and the algorithmic outputs. In the context of AVs, group fairness can be used to ensure that the underlying AI algorithms are independent of these demographic characteristics. As one example, consider a vision-based pedestrian detection system [12], where an AI model is responsible for detecting pedestrians for an AV. Group fairness would require that this system should always make independent decisions regardless of the gender, race, age, ability, and complexion of the pedestrians. One of the most popular notions in the group fairness area is statistical parity, which means the proportions of positive decisions for the marginal group and the majority group should be similar [21]. The metrics developed in the language of statistical parity include equalized odds, risk difference, risk ratio, and odds ratio.

In contrast to group fairness, individual fairness refers to the fact that similar individuals should have similar results [26]. Various studies have defined measures to quantify individual fairness. For instance, Luong et al. [17] developed a situation-testing-based framework to identify similar individuals. In the AV pedestrian detection example mentioned earlier, individual fairness would be achieved if the algorithm yields consistent results for two pedestrians, where all the factors related to the incidents, including approaching direction, speed, lighting condition, etc. are similar— other than the fact that the two pedestrians come from two different demographic groups.

Among other notions of fairness, researchers have developed causal fairness [28] and counterfactual fairness [15] to distinguish spurious disparities from bias. Note that "disparity" is not equivalent to "bias." Only the causal connections between the sensitive information and the algorithmic decisions are considered to be bias. Again, taking the AV pedestrian detection example, causal fairness implies that the race of the pedestrians is not the cause of algorithmic decisions. On the other hand, counterfactual fairness implies that the pedestrian detection system's response should be the same for a pedestrian in the real world and in a counterfactual world where the pedestrian belongs to a different demographic.

Note that none of the fairness metrics mentioned above fits for all situations. Often, being fair in one aspect may result in being biased or unfair in another. Thus, it is important to utilize multiple fairness metrics or to select "a few good measures" to assess the fairness of a model dedicated to a particular AV application. Any fairness framework developed for AVs must clearly outline how these measures or metrics are specifically applied to various autonomous decision-making applications.

*2.1.3 Data Challenges for Measuring Fairness*

Another challenge related to measuring AV fairness lies in the availability of equitable data. In the U.S., the Biden-Harris administration is working towards advancing equitable data principles to identify discrimination in federal policies and programs [30]. In April 2022, the Equitable Data Working Group published a set of recommendations on advancing equitable data for the federal government, such as promoting the use of disaggregated demographic data to benefit underserved communities, leveraging underutilized federal data by allowing inter-agency data sharing, and collaborating with different levels of government, diverse communities, experts, and stakeholders [11]. Such measures are essential to help remove the barriers to achieving equitable data for AV applications.

## 2.2 Equity in AVs

Equity refers to the distribution of opportunity in a way that favors equal outcomes for all [12]. For example, let us consider an autonomous public transit service that follows a route within a small town, "A," twice a day, providing equal transit service or opportunity for all who live there. Now, consider another small town, "B," where the average income of the residents is much lower compared to town A. If we want to start an autonomous public transit service in town B similar to town A, then we may consider having the service operate more than twice a day since the low-income residents living in town B may not be able to own personal vehicles. This treatment is referred to as equity. Similarly, AVs may contribute to equity by providing mobility for those who may not be able or willing to operate vehicles, such as elderly and disabled people.

AV solutions might be expensive enough to cause AV prices to reach a level that is not affordable for lower-income people. How do we address this economic issue? To provide mobility for all, we may wish to focus on maximizing mobility per person rather than maximizing profit. One option might be provisioning a fleet of government-owned, or at least subsidized, shared AVs that facilitate mobility for underrepresented and underserved populations [16,18]. Public regulation may be necessary to require private transportation network companies to provide sufficient service to rural, lower-income, and minority areas; the need for this is shown by the fact that drivers for companies such as Uber and Lyft have shown bias in selecting customers [3]. Autonomous public transit vehicles can offer an equitable solution; for example, by realizing cost savings, the agencies can then reinvest in improving service, which is important since transit riders (particularly bus riders) are disproportionately low-income [1,4,5].

Also, socially responsible AV shared mobility and public transit services may prove to be more beneficial than owning a non-autonomous vehicle, which would justify shared AVs' extra cost. For example, a recent economic assessment of Singapore's autonomous electric microtransit vehicles [19] showed that the total cost of ownership could be reduced by 70% while using autonomous electric microtransit vehicles compared to other microtransit vehicles due to improved accessibility and convenience.

## 2.3 Transparency in AVs

Much progress has been made in the field of data privacy, which is essential since wirelessly connected AVs will generate oceans of data that may, in the wrong hands, be used in discriminatory ways. In Europe, fairness is intertwined with transparency in processing individual user data, as outlined in the General Data Protection Regulation (GDPR). This states that individual data must be "processed lawfully, fairly, and in a transparent manner in relation to the data subject" [9]. However, the GDPR contains technology-neutral language, which may be considered insufficient to ensure transparency. For example, advertising companies are still able to indirectly collect a massive amount of consumer data by tracking consumers' online activities (e.g., "Like" activity on Facebook), although users might assume that the GDPR is ensuring their privacy. Van de Waerdt [25] attributes this non-transparency to "unclear terminology and a lack of interpretational guidance" in the GDPR. A similar situation can happen with AVs, where a plethora of sensors are collecting real-time data about users' mobility preferences which can be shared, inadvertently or intentionally, with third parties in the absence of AV-specific regulation. Transparency of AV-collected data usage is critical for preserving privacy, and a user should have the option of opting out of having their personal information harvested.

The GDPR legislation and related transparency research also have implications for the AI models used in AV technology. The popular AI models currently in use are still considered black-box systems. Considerable effort is being invested in making the models explainable. However, Felzmann et al. [6] argue that explainable or interpretable AI models still lack maturity as they do not consider various users' diverse expectations, implementation concerns, and organizational and societal contexts. Thus, considerations of AV technology users' individual characteristics and contextual factors affecting the use of AVs needs to be embedded in the AI models.

## 3 SOCIAL RESPONSIBILITY FRAMEWORK FOR AVS

Khargonekar and Sampath [13] have proposed an ethical framework for the four steps in a technology development process. These steps include conducting basic research, creating early-stage prototypes, developing initial technology to explore societal impacts, and ultimately producing mature end-products with distinct impacts. They have identified AVs as being in the initial technology phase, where societal impacts are yet to be explored.

Unfortunately, to this point, AV research and development does not include design, implementation, testing, and evaluation criteria explicitly tailored to address the socially responsible operation of AVs, and the larger Intelligent Transportation Systems' ecosystem in which they will be embedded. Worldwide, the advancement of AV technology has outpaced the development of a regulatory framework for it. When AV regulation is on the table, legislative bodies in the U.S., Europe, and Asia have primarily focused on safety, privacy, cybersecurity, and legal liability issues [24]. In the U.S., those states that have implemented AV legislation tend to focus on the definition of an AV, and on protocols and guidelines for AV operations and testing [8]. These regulations almost never consider social responsibility as part of the calculus.

Consequently, to ensure that social responsibility is intrinsic to the deployment of AVs, the development of the technology must move beyond considering only criteria such as safety, traffic delay, and energy efficiency. Instead, it must evolve within a framework that explicitly includes societal impacts such as accessibility, affordability, inclusiveness, equitability, and transparency. Going beyond engineers and computer scientists, experts in the realms of planning, economics, sociology, psychology, and even philosophy need to have active roles to ensure the social responsibility of AVs.

## 4 STRATEGIC REFORM OF HUMAN LABOR

Society must prepare to deal with some detrimental impacts of vehicle automation on lower-skill workers. How can we overcome these? Lessons can be learned from the agricultural system, where past experience shows that despite huge increases in productivity due to technological advances, there is evidence of unwanted social, cultural, political, economic, and geographic consequences due to technological push factors [10]. The social adversities of such technology-driven pushes include a decline in farmers' freedom, the reduction of agricultural employment, and a situation where privately regulated global markets and retailers have gained more power over the farmers. To overcome these issues, a new agricultural framework has been proposed by Horlings and Marsden [10] that focuses on the resilient agricultural practices of the local community. Such a community-focused, adaptive reformation is needed for AV technology. One option is to keep drivers in control of commercial and transit vehicles in case of a challenging situation, as is the case with airplanes and the operation of their autopilot systems. Over the long haul, a more promising strategy would involve reskilling and redeploying workers. In terms of the transportation industry itself, at the same time that AVs make some jobs redundant, they will create others. For example, labor will be needed to build and maintain the supporting computerized infrastructure for AVs, and more jobs in other areas of logistics will be generated when trucking becomes faster and more economical. It is essential that government be at the forefront of helping workers to upskill, adapt, and thrive in this new market.

## 5 FUTURE AVS: SEGREGATION OR UNIFICATION?

True driverless AVs have proven elusive to date, but given the ongoing evolution of the technology, AV adoption seems to be a question of "when" and not "if." When we attain the ability to create autopilot systems that drive door-to-door without human supervision, a plethora of benefits will be unlocked. It is essential that all our citizens enjoy the enhanced safety, traveler ease and comfort, and travel speeds that AVs will foster. We argue that the AV technological push needs to intentionally integrate social responsibility-oriented considerations. Thus, elected officials, policymakers, engineers, computer scientists, planners, social scientists, ethicists, standardization organizations, vehicle manufacturers, and others need to collaborate today to make social responsibility a priority within the rapidly emerging AV ecosystem.

## ACKNOWLEDGMENTS

The authors would like to acknowledge Pamela McKie Foster (Retired Civil Rights Officer of Federal Highway Administration - South Carolina Division) and Dr. Yongkai Wu (Assistant Professor of Electrical and Computer Engineering at Clemson University) for their input to the paper.

## REFERENCES

[1] Ransford A. Acheampong, Federico Cugurullo, Maxime Gueriau, and Ivana Dusparic. 2021. Can autonomous vehicles enable sustainable mobility in future cities? Insights and policy challenges from user preferences over different urban transport options. *Cities* 112, (May 2021), 103134. DOI:https://doi.org/10.1016/j.cities.2021.103134

[2] Henrik Arnelid, Edvin Listo Zec, and Nasser Mohammadiha. 2019. Recurrent Conditional Generative Adversarial Networks for Autonomous Driving Sensor Modelling. In *2019 IEEE Intelligent Transportation Systems Conference (ITSC)*, 1613–1618. DOI:https://doi.org/10.1109/ITSC.2019.8916999

[3] Ryan Calo and Alex Rosenblat. 2017. The taking economy: Uber, information, and power. *Colum. L. Rev.* 117, (2017), 1623.

[4] Kaylla Cantilina, Shanna R Daly, Matthew P Reed, and Robert C Hampshire. 2021. Approaches and Barriers to Addressing Equity in Transportation: Experiences of Transportation Practitioners. *Transportation Research Record* 2675, 10 (2021), 972–985. DOI:https://doi.org/10.1177/0361198121101453

[5] Daniel J. Fagnant and Kara Kockelman. 2015. Preparing a nation for autonomous vehicles: opportunities, barriers and policy recommendations. *Transportation Research Part A: Policy and Practice* 77, (July 2015), 167–181. DOI:https://doi.org/10.1016/j.tra.2015.04.003

[6] Heike Felzmann, Eduard Fosch Villaronga, Christoph Lutz, and Aurelia Tamò-Larrieux. 2019. Transparency you can trust: Transparency requirements for artificial intelligence between legal norms and contextual concerns. *Big Data & Society* 6, 1 (2019), 2053951719860542.

[7] Stefan Feuerriegel, Mateusz Dolata, and Gerhard Schwabe. 2020. Fair AI. *Bus Inf Syst Eng* 62, 4 (August 2020), 379–384. DOI:https://doi.org/10.1007/s12599-020-00650-3

[8] Bryan Gibson and University of Kentucky Transportation Center. 2017. *Analysis of autonomous vehicle policies*. Retrieved April 20, 2023 from https://rosap.ntl.bts.gov/view/dot/32729

[9] Elena Gil González and Paul de Hert. 2019. Understanding the legal provisions that allow processing and profiling of personal data—an analysis of GDPR provisions and principles. *ERA Forum* 19, 4 (April 2019), 597–621. DOI:https://doi.org/10.1007/s12027-018-0546-z

[10] L. G. Horlings and T. K. Marsden. 2011. Towards the real green revolution? Exploring the conceptual dimensions of a new ecological modernisation of agriculture that could 'feed the world.' *Global Environmental Change* 21, 2 (May 2011), 441–452. DOI:https://doi.org/10.1016/j.gloenvcha.2011.01.004

[11] The White House. A Vision for Equitable Data: Recommendations from the Equitable Data Working Group. Retrieved from https://www.whitehouse.gov/wp-content/uploads/2022/04/eo13985-vision-for-equitable-data.pdf

[12] Mhafuzul Islam, Mizanur Rahman, Mashrur Chowdhury, Gurcan Comert, Eshaa Deepak Sood, and Amy Apon. 2020. Vision-Based Personal Safety Messages (PSMs) Generation for Connected Vehicles. *IEEE Transactions on Vehicular Technology* 69, 9 (September 2020), 9402–9416. DOI:https://doi.org/10.1109/TVT.2020.2982189

[13] Pramod P. Khargonekar and Meera Sampath. 2020. A Framework for Ethics in Cyber-Physical-Human Systems∗∗Supported by the University of California, Irvine and the State University of New York, Albany. *IFAC-PapersOnLine* 53, 2 (January 2020), 17008–17015. DOI:https://doi.org/10.1016/j.ifacol.2020.12.1251

[14] Alex Kuefler, Jeremy Morton, Tim Wheeler, and Mykel Kochenderfer. 2017. Imitating driver behavior with generative adversarial networks. In *2017 IEEE Intelligent Vehicles Symposium (IV)*, 204–211. DOI:https://doi.org/10.1109/IVS.2017.7995721

[15] Matt J Kusner, Joshua Loftus, Chris Russell, and Ricardo Silva. 2017. Counterfactual Fairness. In *Advances in Neural Information Processing Systems*, Curran Associates, Inc. Retrieved from https://proceedings.neurips.cc/paper_files/paper/2017/file/a486cd07e4ac3d270571622f4f316ec5-Paper.pdf

[16] Todd Litman. 2020. Autonomous Vehicle Implementation Predictions: Implications for Transport Planning. (January 2020). Retrieved April 20, 2023 from https://trid.trb.org/View/1678741

[17] Binh Thanh Luong, Salvatore Ruggieri, and Franco Turini. 2011. k-NN as an implementation of situation testing for discrimination discovery and prevention. In *Proceedings of the 17th ACM SIGKDD international conference on Knowledge discovery and data mining* (KDD '11), Association for Computing Machinery, New York, NY, USA, 502–510. DOI:https://doi.org/10.1145/2020408.2020488

[18] George Martin. 2019. An Ecosocial Frame for Autonomous Vehicles. *Capitalism Nature Socialism* 30, 4 (October 2019), 55–70. DOI:https://doi.org/10.1080/10455752.2018.1510531

[19] Aybike Ongel, Erik Loewer, Felix Roemer, Ganesh Sethuraman, Fengqi Chang, and Markus Lienkamp. 2019. Economic Assessment of Autonomous Electric Microtransit Vehicles. *Sustainability* 11, 3 (January 2019), 648. DOI:https://doi.org/10.3390/su11030648

[20] Yunpeng Pan, Ching-An Cheng, Kamil Saigol, Keuntaek Lee, Xinyan Yan, Evangelos Theodorou, and Byron Boots. 2019. Agile Autonomous Driving using End-to-End Deep Imitation Learning. DOI:https://doi.org/10.48550/arXiv.1709.07174

[21] Andrea Romei and Salvatore Ruggieri. 2014. A multidisciplinary survey on discrimination analysis. *The Knowledge Engineering Review* 29, 5 (November 2014), 582–638. DOI:https://doi.org/10.1017/S0269888913000039

[22] Debaditya Roy, Tetsuhiro Ishizaka, C. Krishna Mohan, and Atsushi Fukuda. 2019. Vehicle Trajectory Prediction at Intersections using Interaction based Generative Adversarial Networks. In *2019 IEEE Intelligent Transportation Systems Conference (ITSC)*, 2318–2323. DOI:https://doi.org/10.1109/ITSC.2019.8916927

[23] Katja Schwarz, Yiyi Liao, and Andreas Geiger. 2021. On the frequency bias of generative models. *Advances in Neural Information Processing Systems* 34, (2021), 18126–18136.

[24] Araz Taeihagh and Hazel Si Min Lim. 2019. Governing autonomous vehicles: emerging responses for safety, liability, privacy, cybersecurity, and industry risks. *Transport Reviews* 39, 1 (January 2019), 103–128. DOI:https://doi.org/10.1080/01441647.2018.1494640

[25] Peter J. van de Waerdt. 2020. Information asymmetries: recognizing the limits of the GDPR on the data-driven market. *Computer Law & Security Review* 38, (September 2020), 105436. DOI:https://doi.org/10.1016/j.clsr.2020.105436

[26] Rich Zemel, Yu Wu, Kevin Swersky, Toni Pitassi, and Cynthia Dwork. 2013. Learning Fair Representations. In *Proceedings of the 30th International Conference on Machine Learning*, PMLR, 325–333. Retrieved May 6, 2023 from https://proceedings.mlr.press/v28/zemel13.html

[27] Jiakai Zhang and Kyunghyun Cho. 2016. Query-Efficient Imitation Learning for End-to-End Autonomous Driving. DOI:https://doi.org/10.48550/arXiv.1605.06450

[28] Lu Zhang, Yongkai Wu, and Xintao Wu. 2016. A causal framework for discovering and removing direct and indirect discrimination. DOI:https://doi.org/10.48550/arXiv.1611.07509

[29] Shengjia Zhao, Hongyu Ren, Arianna Yuan, Jiaming Song, Noah Goodman, and Stefano Ermon. 2018. Bias and generalization in deep generative models: An empirical study. *Advances in Neural Information Processing Systems* 31, (2018).

[30] 2023. FACT SHEET: White House Office of Science and Technology Policy Announces Progress On Advancing Equitable Data | OSTP. *The White House*. Retrieved April 19, 2023 from https://www.whitehouse.gov/ostp/news-updates/2023/03/24/fact-sheet-white-house-office-of-science-and-technology-policy-announces-progress-on-advancing-equitable-data/

[31] Is ChatGPT the Next AI Milestone for Autonomous Vehicles? - Blog. *VicOne*. Retrieved May 15, 2023 from https://vicone.com/blog/is-chatgpt-the-next-ai-milestone-for-autonomous-vehicles

[32] Illustrating Reinforcement Learning from Human Feedback (RLHF). Retrieved May 15, 2023 from https://huggingface.co/blog/rlhf